\documentclass[twocolumn,superscriptaddress,showpacs,preprintnumbers,prl]{revtex4} 

\usepackage{graphicx}
\usepackage{amsmath}
\usepackage{times}
\usepackage{epsfig}
\usepackage{color}
\usepackage{ulem}   
\begin{document}

\def\salto{\vskip 1cm} \def\lag{\langle} \def\rag{\rangle}
\newcommand{\redit}[1]{\textcolor{red}{#1}}
\newcommand{\blueit}[1]{\textcolor{blue}{#1}}
\newcommand{\magit}[1]{\textcolor{magenta}{#1}}

\newcommand{\MSTD} {Materials Science and Technology Division, Oak Ridge
National Laboratory, Oak Ridge, TN 37831, USA}
\newcommand{\CNMS} {Center for Nanophase Materials Sciences, Oak Ridge National
 Laboratory, Oak Ridge, TN 37831, USA}
\newcommand{\LLNL} {Lawrence Livermore National Laboratory, Livermore, CA
94550, USA}

\title{Systematic reduction of sign errors in many-body calculations
  of atoms and molecules}

\author{Michal Bajdich}           \affiliation {\MSTD}
\author{Murilo L. Tiago}          \affiliation {\MSTD}
\author{Randolph Q. Hood}           \affiliation {\LLNL}
\author{Paul R. C. Kent}        \affiliation {\CNMS}
\author{Fernando A. Reboredo}       \affiliation {\MSTD}
  
\begin{abstract}
  The self-healing diffusion Monte Carlo algorithm (SHDMC) [Phys. Rev.
  B {\bf 79}, 195117 (2009), {\it ibid.} {\bf 80}, 125110 (2009)] is
  shown to be an accurate and robust method for calculating the
  ground-state of atoms and molecules.  By direct comparison with
  accurate configuration interaction results for the oxygen
  atom we show that SHDMC converges systematically towards the
  ground-state wave function.  We present results for the challenging
  N$_2$ molecule, where the binding energies obtained via both energy
  minimization and SHDMC are near chemical accuracy (1 kcal/mol).
  Moreover, we demonstrate that SHDMC is robust enough to find
  the nodal surface for systems at least as large as C$_{20}$ starting
  from random coefficients.  SHDMC is a linear-scaling method, in the
  degrees of freedom of the nodes, that systematically reduces the
  fermion sign problem.
\end{abstract}
\date{\today}

\maketitle
 
Since electrons are fermions, their many-body wave functions must
change sign when the coordinates of any pair are interchanged.  In
contrast, the sign of a bosonic wave functions is unchanged for any
coordinate interchange.  Due to this  misleadingly  small difference, the
ground-state energy of bosons can be determined by quantum Monte Carlo
(QMC) methods \cite{HLRbook,mfoulkesrmp2001} with an accuracy limited
only by computing time, while QMC calculations of fermions are either
exponentially difficult, or are stabilized by imposing a systematic
error, a direct consequence of our lack of knowledge of the fermionic
nodal surface.  Therefore, one of the most important problems in
many-body electronic structure theory is to accurately find
representations of the fermion nodes~\cite{ceperley91,mtroyerprl2005},
the locations where the fermionic wave function changes sign, the
so-called ``fermion sign problem''.

The sign problem 
  limits (i) the number of physical systems where {\it ab
    initio} QMC can be applied and (ii) our ability to improve
  approximations of density functional theory (DFT) using QMC
  results~\cite{ceperley80}. More importantly, it limits our overall
  understanding of the effects of interactions in fermionic systems. Therefore, a method
  to circumvent the sign problem with reduced computational cost could
  transform Condensed Matter Theory, Quantum Chemistry and Nuclear
  Physics among other fields.

Arguably the most accurate technique for calculating the ground-state
of a many-body system with more than $20$ fermions is
diffusion Monte Carlo (DMC).  
The standard DMC
algorithm~\cite{ceperley80} finds the lowest energy of all wave functions
that share the nodal surface $S_{T}({\bf R})$ imposed by a trial wave
function $\Psi_{T}({\bf R})$.  This is the fixed-node approximation
where the resultant energy $E_{DMC}$ is a rigorous upper bound of the
exact ground-state energy \cite{anderson79,reynolds82}. The exact
ground-state energy is obtained only when $\Psi_{T}({\bf R})$ has the
same nodal surface as the exact ground-state wave function.  

If the exact nodes are not provided, the implicit fixed-node
ground-state wave function $\Psi_{FN}({\bf R})$ will exhibit
discontinuities in its gradient~\cite{reynolds82,keystone}
(i.e. kinks) on some parts of $S_{T}({\bf R})$. We recently
proved~\cite{keystone} that by locally smoothing these discontinuities
in $\Psi_{FN}({\bf R})$, a new trial wave function can be obtained
with improved nodes. This proof enables an algorithm that
systematically moves the nodal surface  $S_{T}({\bf R})$  
towards the one of an eigen-state.  If the form of
trial wave function is sufficiently flexible, and given sufficient
statistics, this process leads to an exact eigen-state wave
function~\cite{keystone,rockandroll}.  We named the method
self-healing DMC (SHDMC), since the trial wave function is
self-corrected in DMC and can recover even from a poor starting
point.
 
In this Letter, we report the first applications of SHDMC to real
atoms and molecules (O, N$_2$, C$_{20}$).  SHDMC energies are
  within error bars of DMC calculations using the current state of the
  art approach~\cite{umrigar05,umrigar07}. Tests of SHDMC for
C$_{20}$ demonstrate that our method can be applied at the
nanoscale. Its cost scales linearly with the number of independent
degrees of freedom of the nodes with an accuracy limited only by the
achievable statistics and choice of representation of the nodes.

{\it Brief review of SHDMC} --- SHDMC is fundamentally different from
optimization methods used in variational Monte Carlo (VMC):
\cite{HLRbook,mfoulkesrmp2001} (i) the wave function is directly
optimized based on a property of the nodal surface and not on the
local energy or its variance, and (ii) the nodes are optimized at the
DMC level (as opposed to a VMC based algorithm).

Using a short-time many-body propagator, SHDMC samples the
coefficients of an improved wave function removing the artificial
derivative discontinuities of $\Psi_{FN}({\bf R})$ arising from the
inexact nodes.  Repeated application of this method results in the
best nodal surface for a given basis.  For wave functions expanded in
a complete basis it can be shown that the final accuracy is limited
only by the statistics~\cite{keystone,rockandroll}. 

In SHDMC (see Refs.~\onlinecite{keystone,rockandroll} for details),
the weighted walker distribution is~\cite{ceperley80}
\begin{align}\label{eq:evoltau}
  f({\bf R},\tau^\prime+\tau) &= \Psi_T^*({\bf R},\tau^{\prime}) \left[ e^{-\tau (\hat{{\mathcal H}}_{FN}-E_T) } 
\Psi_T({\bf R},\tau^\prime) \right] \\ \nonumber
                & =  \lim_{N_c \rightarrow \infty} \frac{1}{N_c} \sum_{i=1}^{N_c} W_i^j(k) \delta \left({\bf R-R}_i^j \right), 
\end{align}
where
\begin{align}
\label{eq:trialwf}
\Psi_T({\bf R},\tau^{\prime}) = 
 e^{J({\bf R})} 
\sum_n^{\sim} \lambda_n(\tau^\prime)  \Phi_n({\bf R})
\end{align}
is a trial function where $\sum_n^{\sim}$ represents a truncated sum, 
$\{\Phi_n({\bf R})\}$ forms a complete
orthonormal basis of the antisymmetric Hilbert space 
and $e^{J({\bf R})}$ is a  symmetric Jastrow
factor.  In Eq.~(\ref{eq:evoltau}),
$\hat{{\mathcal H}}_{FN}$ is the fixed-node
Hamiltonian [$ \hat{{\mathcal H}}_{FN} $ is the
many-body Hamiltonian with an infinite potential at the
nodes of $\Psi_T({\bf R},\tau^{\prime}) $] and $E_T$ is an energy
reference.  Next, ${\bf R}_i^j$ corresponds to the position of the
walker $i$ at step $j$ of $N_c$ equilibrated configurations. The
weights $W_i^j(k)$ are given by
\begin{align}
\label{eq:weights}
W_i^j(k)\!=\!e^{-\left[
E_i^j(k)-E_{T}
\right] \tau}
\text{with }
  E_i^j(k)\! = \!\frac{1}{k}\!\sum_{\ell=0}^{k-1}\! E_L({\bf R}_i^{j-\ell}),
\end{align} 
where $E_T$ in Eq.~(\ref{eq:weights}) is periodically adjusted so that
$\sum_i W_i^j(k)\approx N_c$ and $\tau$ is $k\delta\tau$ (with $k$
being a number of steps and $\delta\tau$ a standard DMC time step).

From Eq.~(\ref{eq:evoltau}), one can formally obtain
\begin{align}
\label{eq:nextevol}
 \tilde \Psi_T({\bf R},\tau^{\prime}+\tau)  =   f({\bf R},\tau^{\prime}+\tau)/ 
 \Psi_T^*({\bf R},\tau^{\prime}).
\end{align}
We now define the local smoothing function to be 
\begin{align}
\label{eq:deltaexp}
\tilde \delta \left( {\bf R^{\prime},R} \right) =
\sum_n^{\sim}
e^{J({\bf R^{\prime}})} \Phi_n({\bf R^{\prime}}) \Phi_n^*({\bf R}) e^{-J({\bf R})} .
\end{align}

Applying Eq.~(\ref{eq:deltaexp}) to both sides of Eq.~(\ref{eq:nextevol}), 
using Eq.~(\ref{eq:evoltau}),
and integrating over ${\bf R}$ 
we obtain
\begin{align}
\label{eq:sequence}
\Psi_T({\bf R},\tau^{\prime}+\tau) 
= e^{J({\bf R})} 
\sum_n^\sim \lambda_n(\tau^\prime+\tau)  \Phi_n({\bf R}),
\end{align}
with 
\begin{align}
\label{eq:lambda}
\lambda_n(\tau^\prime\!+\!\tau) =\!\!\lim_{N_c \rightarrow \infty} \frac{1}{\mathcal{N}} \sum_i^{N_c} 
W_i^j(k)  e^{-2J({\bf R}_i^j)} \frac {\Phi_n^* ({\bf R}_i^j)}
 { \Phi_T^* ({\bf R}_i^j,\tau^\prime)}
\end{align}
where $ \mathcal{N}= \sum_{i=1}^{N_c} e^{-2J({\bf R}_i^j)} $
normalizes the Jastrow factor.  These new
$\lambda_n(\tau^\prime+\tau)$ [Eq. (\ref{eq:lambda})] are used to
construct a new trial wave function [Eq.~(\ref{eq:trialwf})]
recursively within DMC (therefore the name self-healing DMC). The
weights in Eq. (\ref{eq:weights}) can be evaluated within (i) a
branching algorithm~\cite{keystone} for $\tau^\prime \rightarrow
\infty$ or (ii) a fixed population scheme for small $\tau^\prime
$~\cite{rockandroll,umrigar_private}.  The former method is more
robust, but the latter improves final convergence.  Equation
(\ref{eq:lambda}) can be related to the maximum-overlap method used
for bosonic wave functions~\cite{reatto82}.

Since SHDMC is targeted for large systems we report validations
using pseudopotentials.


{\it Validation of SHDMC with configuration interaction (CI) calculations for the O atom} 
--- In short, CI is the
diagonalization of the many-body Hamiltonian in a truncated basis of
Slater determinants.
We chose to study the $^3$P ground-state of the O atom because it has at least two
valence electrons in both spin channels~\cite{burkatzki07}.
The single-particle orbitals were expanded in VTZ and V5Z Gaussian basis sets~\cite{burkatzki07}
using the GAMESS~\cite{games09} code.  To
facilitate a direct comparison between SHDMC and CI, no
Jastrow factor was employed. 



\begin{figure}
\includegraphics[width=0.95\linewidth,clip=true]{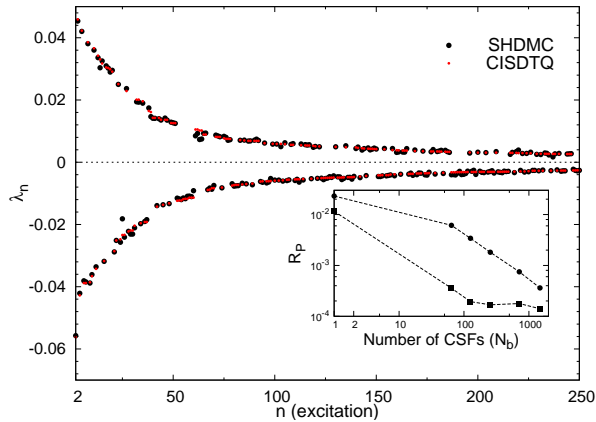}
\caption{Comparison of the values of the coefficients $\lambda_n$
  corresponding to the first 250 excitations of a
  converged SHDMC trial wave function (large black circles) with a large
  CISDTQ wave function (small red circles) for the oxygen atom. 
  The first coefficient of the expansion, 0.9769, is
  not shown.  Inset: Residual projection ($R_P=1-|\langle\Psi_{\rm SHDMC}|\Psi_{\rm CI}\rangle /\langle
\Psi_{\rm CI}|\Psi_{\rm CI}\rangle|$) as a function of the number of
  CSFs included: circles $R_P$ obtained with the full
  CISDTQ norm, squares $R_P$ obtained with the truncated CISDTQ norm.
\label{fg:O-SHDMC-CI}}
\end{figure}

Figure~\ref{fg:O-SHDMC-CI} shows a direct comparison of the first 250
converged coefficients $\lambda_n$ obtained using SHDMC
with those from the largest CI calculation (see Table~\ref{Otable}). 
The initial SHDMC trial wave function was the Hartree--Fock (HF) solution, and the final SHDMC
coefficients resulted from sampling the 1481 most significant
excitations in the  CI. We used 
$\delta\tau=0.01\,a.u.$, $\tau=0.5\,a.u.$, and $16$
iterations of trial wave function projection 
( $\approx 6\times10^7$ sampled configurations).

\begin{table}
 \caption{Total energies (and correlation \% in \{\}) for the ground-state of O obtained with CI, 
    coupled-cluster (CCSD(T)~\cite{ccsdt}) and SHDMC (no Jastrow). Other symbols defined in the text. 
  }
\label{Otable}
\begin{ruledtabular}

\begin{tabular}{l c l c l }
  &  \multicolumn{2}{c}{VTZ} & \multicolumn{2}{c}{V5Z} \\
  \cline{2-3} \cline{4-5}
  Method & $N_b$ & E [Ha]\{[\%]\} & $N_b$ & E [Ha]\{[\%]\} \\
\hline
 CI\footnotemark[1] & 775182 &-15.88258\{89.0\} &1762377 &-15.89557\{95.7\} \\
 CCSD(T)\footnotemark[2] & - &-15.88204\{88.8\} & -      &-15.90166\{98.8\}\\
 SHDMC     &539& -15.9003(2)\{98.1(1)\} &1481&-15.9040(4)\{100.0(2)\}\\
\end{tabular}

\end{ruledtabular}
\footnotetext[1]{full-CI in VTZ and CISDTQ in V5Z.}
\footnotetext[2]{from Ref.~\cite{burkatzki07}.}

\end{table}

Figure ~\ref{fg:O-SHDMC-CI} shows the
excellent agreement between the coefficients $\lambda_n$ obtained 
independently by SHDMC and CI. A perfect agreement is
guaranteed only in the limit of a complete basis and $N_c \rightarrow
\infty$. The small differences in Fig.~\ref{fg:O-SHDMC-CI} are due to
the truncation of the expansion and the stochastic error in
$\lambda_n$.
The inset shows the residual projection
as a function of the total number $N_b$
of CSFs included in the expansion, normalized either using the entire CI
expansion (circles) or using a $\Psi_{\rm CI}$ that included
only the $\lambda_n$ sampled in SHDMC (squares). The residual projection is much smaller
for the truncated norm than the full norm illustrating
that most of the error in $\Psi_{\rm SHDMC}$ is from truncation and
not limited statistics. Similar results were obtained for the C atom
(not shown).

{\it Validation with Energy Minimization for N$_2$ }--- 
We also compared the VMC and DMC energies of wave functions optimized
with energy minimization in VMC (EMVMC)~\cite{umrigar05,umrigar07} and
SHDMC using the QWALK~\cite{wagner09} code.  EMVMC can be briefly
described as a generalized CI with an additional Jastrow factor
(sampling the Hamiltonian stochastically and solving a
generalized eigenvalue problem).  Several bases were obtained from
series of complete active space (CAS) and restricted active space
(RAS)~\cite{ras} multiconfiguration self-consistent field (MCSCF)
calculations [distributing 10 electrons into $m$ active orbitals:
CAS(10,$m$)]. We retained the $N_b$ basis functions with
coefficients of absolute value larger than a given cutoff.
Subsequently, for each basis, we performed energy minimization of the
Jastrow and the coefficients of trial wave function using a mixture of
95\% of energy and 5\% of variance.  We also sampled these $N_b$
coefficients in SHDMC recursively starting from HF solution. For a clear comparison we used the
same Jastrow in EMVMC and SHDMC.
 
\begin{figure}
\includegraphics[width=1.00\linewidth,clip=true]{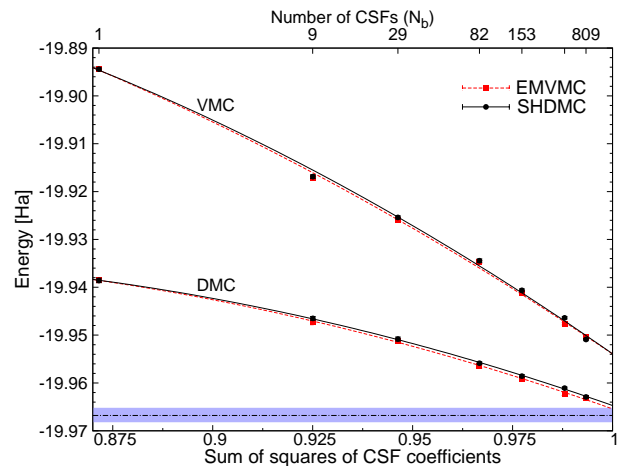}
\caption{Total energies 
  obtained for N$_2$ with VMC and DMC methods for wave functions optimized via
  EMVMC~\cite{umrigar05} (squares) and SHDMC (circles)
  as a function of the square of the norm of the CI coefficients
  retained in the basis [$\sum_n (\lambda^{\rm MCSCF}_n)^2 $].
  The lines are parabolic extrapolations to 1.
  The dot-dashed line represents the scalar relativistic
  core-corrected estimate of the exact energy (see
  Table~\ref{n2table}). The shaded area is the region of
  chemical accuracy.
  \label{fg:N2-SHDMC-EM}}
\end{figure}

We performed these calculations for the ground-state ($^1\Sigma^+_g$) of N$_2$ at
the experimental geometry~\cite{Ruscic}. 
Figure~\ref{fg:N2-SHDMC-EM} shows the resulting VMC and DMC energies
obtained for wave functions optimized independently with EMVMC and SHDMC
methods for the largest RAS(10,43) (2629447 CSFs yielding E=-19.921717) 
Slater-Jastrow wave function (See also Table~\ref{n2table}). 
In EMVMC , as previously observed for C$_2$ and
Si$_2$~\cite{umrigar07}, we found a systematic reduction in the fixed-node errors, even when
starting from the smallest CAS wave function (see
Table~\ref{n2table}).  When we compare with SHDMC optimized wave
functions we find an excellent agreement in both VMC and DMC
energies. Therefore, SHDMC improves the nodes systematically starting
from the HF ground-state.

Since retaining all the determinants in the wave function would be
costly, we performed calculations 
with different $N_b$ to extrapolate (quadratically) the final
energies as $\sum_n (\lambda^{\rm MCSCF}_n)^2 \to 1$ (see  Fig.~\ref{fg:N2-SHDMC-EM}).  The
extrapolated DMC energies reached chemical accuracy (see also
 Table~\ref{n2table}).

\begin{table}
  \caption{
    Comparison of total and binding DMC energies of N$_2$ 
    for wave functions optimized with EMVMC and SHDMC for increasingly  
      larger basis (see text). All SHDMC calculations
      started from the single HF determinant.
    Binding energies were obtained using an
    atomic energy$^c$ of  -9.80213(5) Ha,
    a core-correlations correction of 1.4 mHa~\cite{Bytautas}, and a 
    zero point energy of 5.4 mHa~\cite{Ruscic}.
  }
\label{n2table}
\begin{ruledtabular}
\begin{tabular}{l c c c c }
  & \multicolumn{2}{c}{Total energy [Ha]} & \multicolumn{2}{c}{Binding energy [eV]} \\
  \cline{2-3} \cline{4-5}
   Wave function & EMVMC & SHDMC & EMVMC & SHDMC \\
\hline
 1 determinant     &   \multicolumn{2}{c}{-19.9362(5)} &  \multicolumn{2}{c}{9.07(1)} \\
 CAS(10,14) & -19.9536(6)  &  -19.9536(6) & 9.54(2) & 9.54(2) \\
 RAS(10,35) & -19.9639(4)  &  -19.9627(4) & 9.83(1) & 9.79(1) \\
 RAS(10,43) & -19.9654(4)  &  -19.9647(4) & 9.87(1) & 9.85(1) \\
 Estimated exact & \multicolumn{2}{c}{-19.9668(2)\footnotemark[1]} & \multicolumn{2}{c}{-9.900(1)\footnotemark[2]} \\
\end{tabular}
\end{ruledtabular}
\footnotetext[1]{Based on the scalar relativistic core-corrected estimate from 
Ref.~\cite{Bytautas}.}
\footnotetext[2]{Using the experimental value from Ref.~\cite{Ruscic}. } 
\footnotetext[3]{Based on a large multi-determinant DMC calculation.}
\end{table}

{\it Proof of principle in larger systems} --- Figures
\ref{fg:O-SHDMC-CI} and \ref{fg:N2-SHDMC-EM} show that SHDMC produces
reliable and accurate results for small systems starting form the HF nodes.
It is also important to demonstrate that SHDMC is a robust
approach that can find the correct nodal surface topology of much larger
systems even when starting from random nodal surfaces.
 
\begin{figure}
\includegraphics[width=1.00\linewidth,clip=true]{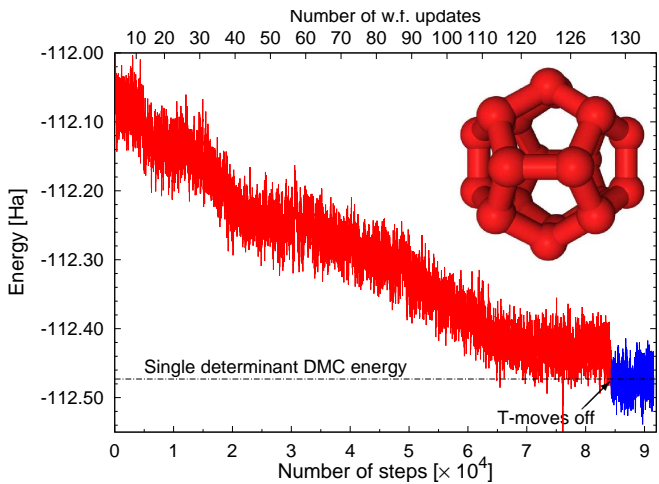}
\caption{Proof of principle of SHDMC for larger systems. 
  Initial evolution of the average local energy for a SHDMC run with
  branching\cite{keystone} generated for C$_{20}^{+2}$, with random
  initial coefficients (see text). 
  Inset: calculated icosahedral cluster C$_{20}^{+2}$.
\label{fg:C20-SHDMC-OLD}}
\end{figure}

Figure~\ref{fg:C20-SHDMC-OLD} shows proof of principle results
obtained for a C$_{20}$ fullerene. 
These calculations used the branching SHDMC  
algorithm[\onlinecite{keystone}] implemented by us in  CASINO~\cite{casino}.
Two electrons were removed from the system to obtain a
non-interacting DFT ground-state wave function invariant under any
transformation belonging to the icosahedral group ($I_h$) symmetry. 
The orbitals were obtained directly with the real
space code PARSEC~\cite{parsec} and classified according to their irreducible
representations for $I_h$ and its subgroup $D_{2h}$. 
For this calculation 694 excitations (determinants) were 
sampled. No CI prefiltering of determinants is required; we only
use the selection rules of both $I_h$ and $D_{2h}$ symmetries.

The C$_{20}^{+2}$ system
has a large DFT gap (5.53 eV) which is often associated with a dominant role of
the non-interacting solution in the many-body wave function. The $\lambda_0$
coefficient is expected to dominate the final optimized trial wave
function.
All initial coefficients $\lambda_n$ of $\Psi_T({\bf R})$ were set to random values, but
for $\lambda_0$ which was set to zero.  New $\lambda_n$ 
values were sampled with $\sim 5094$ walkers every $100$ DMC steps. We found that when the
quality of the wave function is poor, it is better (i) to update
$\lambda_n$ frequently (after only $4$ samplings), and (ii) to use the
T-moves approximation~\cite{casula06} which limits persistent
configurations. As the quality of the wave-function improved, we
gradually increased the accumulation time (up to 96 samplings) and
removed the T-moves approximation (which, in practice, hinders the
final SHDMC convergence).
Figure~\ref{fg:C20-SHDMC-OLD} shows that SHDMC can correct nodal
errors as large as $0.5$ Ha.  The calculation was stopped when we
obtained an energy of $-112.487(2)$ Ha compared with the single
determinant energy of $-112.473(1)$ Ha.
We have confidence that SHDMC can be applied to cases where the
nodal structure of the ground-state is completely unknown since it
is successful and converges to the expected result
starting from random.

The SHDMC recursive runs required $220$ hrs on $1024$ processors (Cray
XT4).  This can be reduced to $\sim 100$ hrs starting from the ground
state determinant.  Comparable EMVMC calculations with the same basis
were unsuccessful, presumably due to the statistical errors in the
Hessian and overlap matrices.  The energy was not improved with EMVMC
($-112.488(3)$ Ha) even selecting a basis with the largest $104$
coefficients of the $694$ sampled in SHDMC. The estimated running time
for EMVMC with CASINO 2.5 using $N_b=694$ and just $400$
configurations~\cite{fn:configurations} on $1024$ processors is
already $\sim 100$ hrs, suggesting that for C$^+_{20}$ SHDMC is faster
than EMVMC. However, both methods can be improved for large $N_b$
(e.g. as in Ref.  \onlinecite{nukala09}), by removing redundant IO etc.

{\it Summary} --- We have shown that the SHDMC wave function converges
to the ground-state of our best CI calculations and is systematically
improved as the number of coefficients sampled increases and the
statistics are improved. SHDMC presents equivalent accuracy to the
EMVMC approach~\cite{umrigar05,umrigar07} starting from random
  coefficients.  SHDMC is numerically robust and can be automated.

The number of independent degrees of freedom of the nodes
  increases exponentially with the number of
  electrons.~\cite{rockandroll}  Since EMVMC is based on VMC, the
prefactor for its computational cost is much smaller than SHDMC.
However, the number of quantities sampled in EMVMC is quadratic with
respect to the number of degrees of freedom.  In addition, EMVMC requires
inverting a noisy matrix. 
These requirements cause EMVMC to
scale at least quadratically. In contrast, SHDMC 
only requires one to sample a number of quantities linear in the
number of optimized degrees of freedom.  Therefore,
a crossover between the methods is expected for systems
of sufficient size or complexity.
Tests on the large C$_{20}^{+2}$ fullerene system demonstrate that
SHDMC is robust and that the nodes are systematically improved even
starting from a random coefficients in the trial wave
function. This shows that SHDMC can be used to find the nodes of
unknown complex systems of unprecedented size.

We thank D. Ceperley, R. M. Martin and C. J. Umrigar for critically
reading the manuscript and useful comments.  This research used
computer resources supported by the U.S. DOE Office of Science under
contract DE-AC02-05CH11231 (NERSC) and DE-AC05-00OR22725
(NCCS). Research sponsored by U.S. DOE BES Division of Materials
Sciences \& Engineering (FAR, MLT) and ORNL LDRD program (MB). The
Center for Nanophase Materials Sciences research was sponsored by the
U. S. DOE Division of Scientific User Facilities (PRCK). Research at
LLNL was performed under U.S. DOE contract DE-AC52-07NA27344 (RQH).


\begin{thebibliography}{99}
\bibitem{ras} We included excitations up to quadruple level.
\bibitem{fn:configurations} These configurations are not enough for $N_b=100$.
\end{thebibliography}
\end{document}